\documentclass[a4paper]{article}
\RequirePackage[english]{babel}
\RequirePackage[latin1]{inputenc}
\RequirePackage[T1]{fontenc}
\RequirePackage{mathrsfs}
\RequirePackage{amsmath}
\RequirePackage{amssymb}
\RequirePackage{amsbsy}
\RequirePackage{bm}
\begin{document}
%%%%%%%%%%%%%%%%%%%%%%%%%%%%%%%%%%%%%%%%%%%%%%%%%%%%%%%%%%%%%%%%%%%%%%%%%%%%%%%%%%%%%%%%%%%%%%%%%%%
\title{\bf{Conformal Standard Model}}
\author{Luca Fabbri\\ 
\footnotesize DIPTEM Sez. Metodi e Modelli Matematici dell'Universit\`{a} di Genova \& \\
\footnotesize INFN Sez. di Bologna and Dipartimento di Fisica dell'Universit\`{a} di Bologna}
\date{}
%%%%%%%%%%%%%%%%%%%%%%%%%%%%%%%%%%%%%%%%%%%%%%%%%%%%%%%%%%%%%%%%%%%%%%%%%%%%%%%%%%%%%%%%%%%%%%%%%%%
\maketitle
%%%%%%%%%%%%%%%%%%%%%%%%%%%%%%%%%%%%%%%%%%%%%%%%%%%%%%%%%%%%%%%%%%%%%%%%%%%%%%%%%%%%%%%%%%%%%%%%%%%
\begin{abstract}
In recent papers we have constructed the conformal theory of metric-torsional gravitation, and in this paper we shall include the gauge fields to study the conformal $U(1)\times SU(2)$ Standard Model; we will show that the metric-torsional degrees of freedom give rise to a potential of conformal-gauge dynamical symmetry breaking: consequences are discussed.
\end{abstract}
%%%%%%%%%%%%%%%%%%%%%%%%%%%%%%%%%%%%%%%%%%%%%%%%%%%%%%%%%%%%%%%%%%%%%%%%%%%%%%%%%%%%%%%%%%%%%%%%%%%
\section*{Introduction}
The reasons for which conformal Weyl gravity is remarkable are two: Weyl gravity possesses solutions that, for the solar system, are able to approximate the Einstein gravity solutions, while, as the scales increase to galaxies, they are also able to fit the rotation curves \cite{m-k,Mannheim:2010ti}, and for the universe in its entirety, they address the cosmological constant problem \cite{m}, so to be able of replacing dark forms of matter and energy \cite{Mannheim:2005bfa}; the equations of Weyl gravity are fourth-order differential equations with dimensionless constants whose renormalizability is useful to address the gravitational quantization \cite{s}. Thus far, the entire discussion has been carried out in terms of conformal Weyl gravity in the purely metric case, about which a general discussion can be found in \cite{Mannheim:2011ds}, but on the other hand one has to keep in mind that in order to build a complete quantum theory not only the metric but also the torsional degrees of freedom have to be considered, because according to the Wigner classification of quantum particles in terms of their mass and spin, matter fields possess not only energy but also spin density, the former being related to the curvature while the latter being linked to the torsion of the spacetime: thus for a conformal metric-torsional gravitation to be constructed one needs to define, beside the conformal transformations for the metric, also the corresponding conformal transformations for the torsion tensor \cite{sh}. Once the conformal properties of the metric as well as torsion are settled, one needs to define a conformally covariant metric-torsional curvature in $(1+3)$-dimensional spacetimes upon which to build the conformal gravitational field theory as it was done in \cite{f}. This theory has further been applied to the special case of the conformal massless Dirac field theory as in \cite{fabbri}.

Then it is important to recall that not only the metric-torsional degrees of freedom but also the gauge degrees of freedom have to be considered because matter fields possess not only energy-spin but also current density: for the conformal metric-torsional gravitation to include gauge fields one should define conformal transformations for the gauge fields, which nonetheless are trivial as it is widely known. Because of this fact we have that the conformally covariant gauge strength in $(1+3)$-dimensions is unchanged with respect to the standard case in which the conformal invariance of gauge field theories was already a character of the model \cite{e-f-p}. The inclusion of scalar fields then follows \cite{mannheim-kazanas}.

However when in the purely metric curvature the torsion is accounted then the metric-torsional curvature changes its properties, and consequently the effects of its coupling to other fields, so that while the background is decoupled from the gauge strength, and therefore such a modification has no influence on the gauge sector, the background has peculiar coupling to the scalar fields, and henceforth this modification greatly affects the scalar fields as we shall show in the present paper. This allows us to merge the results of \cite{F} and \cite{FABBRI} obtaining a metric-torsion conformal $U(1)\times SU(2)$-gauge symmetric interaction for the Dirac and scalar fields to get the conformal Standard Model.
%%%%%%%%%%%%%%%%%%%%%%%%%%%%%%%%%%%%%%%%%%%%%%%%%%%%%%%%%%%%%%%%%%%%%%%%%%%%%%%%%%%%%%%%%%%%%%%%%%%
%%%%%%%%%%%%%%%%%%%%%%%%%%%%%%%%%%%%%%%%%%%%%%%%%%%%%%%%%%%%%%%%%%%%%%%%%%%%%%%%%%%%%%%%%%%%%%%%%%%
\section{Conformal Geometry}
In this paper we shall follow all notation and conventions of \cite{f,fabbri} about conformal gravity. In particular and as usual, the metric is given by $g_{\mu\nu}$ while the connection is given by $\Gamma^{\alpha}_{\rho\sigma}$ whose antisymmetric part in the lower indices is torsion $\Gamma^{\alpha}_{\rho\sigma}-\Gamma^{\alpha}_{\sigma\rho}=Q^{\alpha}_{\phantom{\alpha}\rho\sigma}$ also known as Cartan tensor; in the following of the paper we will assume the metric-compatibility condition for the connection, that is we shall insist on the fact that the metric is the constant for the covariant derivatives $D_{\alpha}g_{\mu\nu}=0$ spelling out that the metric properties and the differential features of the space have characters that are compatible, although because of the presence of torsion, they remain independent, as it is clear from the fact the metric-compatibility condition is equivalent to the decomposition
\begin{eqnarray}
&\Gamma^{\sigma}_{\phantom{\sigma}\rho\alpha}=
\frac{1}{2}g^{\sigma\theta}[Q_{\rho\alpha\theta}+Q_{\alpha\rho\theta}+Q_{\theta\rho\alpha}
+(\partial_{\rho}g_{\alpha\theta}+\partial_{\alpha}g_{\rho\theta}-\partial_{\theta}g_{\rho\alpha})]
\label{connection}
\end{eqnarray}
in terms of both metric and Cartan torsion tensor. An equivalent formalism can be introduced. In it we consider the constant Minkowskian metric $\eta_{ij}$ and a basis of vierbein $e_{\alpha}^{i}$ such that we have the relationship $e_{\alpha}^{p}e_{\nu}^{i}\eta_{pi}=g_{\alpha\nu}$ together with the spin-connection $\omega^{ip}_{\phantom{ip}\alpha}$ for which no torsion tensor can possibly be defined because of the different type of Latin and Greek indices; vierbein-compatibility conditions are imposed accordingly, leading to the conditions $D_{\alpha}e_{\mu}^{j}=0$ and therefore $D_{\alpha}\eta_{ij}=0$ respectively yielding the formula
\begin{eqnarray}
&\omega^{i}_{\phantom{i}p\alpha}=
e^{i}_{\sigma}(\Gamma^{\sigma}_{\rho\alpha}e^{\rho}_{p}+\partial_{\alpha}e^{\sigma}_{p})
\label{spin-connection}
\end{eqnarray} 
with the property $\omega^{ip}_{\phantom{ip}\alpha}=-\omega^{pi}_{\phantom{pi}\alpha}$, showing that it is therefore possible to employ the connection and the vierbein to give rise to the spin-connection, which is antisymmetric in the two Latin indices, and again vierbein and spin-connection are taken independent. The former formalism indicated with Greek letters and the latter formalism indicated with Latin letters are respectively denoted as spacetime formalism and world formalism, and they are equivalent, although in the last formalism there is the advantage for which the introduction of the spinorial structure is possible. To this extent, we introduce the set of $\gamma_{a}$ matrices verifying the Clifford algebra $\{\gamma_{a},\gamma_{b}\} =2\mathbb{I}\eta_{ab}$ from which it is further possible to define the $\sigma_{ab}$ matrices $\sigma_{ab}=\frac{1}{4}[\gamma_{a},\gamma_{b}]$ such that $\{\gamma_{a},\sigma_{bc}\} =i\varepsilon_{abcd} \gamma\gamma^{d}$ which are the generators of the spinorial transformation defining spinor fields $\psi$, whose dynamics defined in terms of the spinor-connection $\Omega_{\rho}=\frac{1}{2}\omega^{ij}_{\phantom{ij}\rho}\sigma_{ij}$ is encoded through the spinor-covariant derivatives $D_{\rho}\psi=\partial_{\rho}\psi+\Omega_{\rho}\psi$ with respect to which the constancy of the $\gamma_{a}$ matrices is automatic. Thus the geometrical background is given, and conformal properties have next to be assigned according to the usual metric conformal transformation as in the following
\begin{eqnarray}
&g_{\alpha\theta}\rightarrow\sigma^{2}g_{\alpha\theta}
\end{eqnarray}
and by defining $Q^{\sigma}_{\sigma\alpha}=Q_{\alpha}$ as the torsion trace vector we postulate the torsional conformal transformations to be given by
\begin{eqnarray}
&Q^{\sigma}_{\phantom{\sigma}\rho\alpha}\rightarrow Q^{\sigma}_{\phantom{\sigma}\rho\alpha}
+q\sigma^{-1}(\delta^{\sigma}_{\rho}\partial_{\alpha}\sigma
-\delta^{\sigma}_{\alpha}\partial_{\rho}\sigma)
\end{eqnarray}
for a given parameter $q$, and as it is clear by taking the contraction it is on the torsion trace alone that the conformal transformation is loaded, thus implying that $Q_{\beta\rho\mu} \varepsilon^{\beta\rho\mu\alpha}=-6V^{\alpha}$ known as torsion dual axial vector and the remaining irreducible part of torsion are conformally covariant; the curvature tensor 
\begin{eqnarray}
&G^{\rho}_{\phantom{\rho}\xi\mu\nu}
=\partial_{\mu}\Gamma^{\rho}_{\xi\nu}-\partial_{\nu}\Gamma^{\rho}_{\xi\mu}
+\Gamma^{\rho}_{\sigma\mu}\Gamma^{\sigma}_{\xi\nu}
-\Gamma^{\rho}_{\sigma\nu}\Gamma^{\sigma}_{\xi\mu}
\end{eqnarray}
or equivalently in the form $G^{\rho}_{\phantom{\rho}\xi\mu\nu}e^{i}_{\rho}e^{\xi}_{j}
=G^{i}_{\phantom{i}j\mu\nu}$ given by
\begin{eqnarray}
&G^{i}_{\phantom{i}j\mu\nu}
=\partial_{\mu}\omega^{i}_{j\nu}-\partial_{\nu}\omega^{i}_{j\mu}
+\omega^{i}_{k\mu}\omega^{k}_{j\nu}-\omega^{i}_{k\nu}\omega^{k}_{j\mu}
\label{curvature}
\end{eqnarray}
contains torsion implicitly through the connection, and from it it is possible to define a modified metric-torsional curvature tensor
\begin{eqnarray}
&M_{\alpha\theta\mu\nu}
=G_{\alpha\theta\mu\nu}+(\frac{1-q}{3q})(Q_{\theta}Q_{\alpha\mu\nu}-Q_{\alpha}Q_{\theta\mu\nu})
\label{modifiedcurvature}
\end{eqnarray}
containing torsion also explicitly, and which has the same symmetries, and it is such that its irreducible decomposition
\begin{eqnarray}
\nonumber
&T_{\alpha\theta\mu\nu}=M_{\alpha\theta\mu\nu}
-\frac{1}{2}(M_{\alpha\mu}g_{\nu\theta}-M_{\theta\mu}g_{\nu\alpha}
-M_{\alpha\nu}g_{\mu\theta}+M_{\theta\nu}g_{\mu\alpha})+\\
&+\frac{1}{6}M(g_{\alpha\mu}g_{\nu\theta}-g_{\theta\mu}g_{\nu\alpha})
\label{conformalcurvature}
\end{eqnarray}
does not only have the same symmetries and is traceless but it is also conformally covariant; and finally, it is in terms of the set of three free parameters $A$, $B$, $C$ that it is possible to define the parametric curvature tensor
\begin{eqnarray}
\nonumber
&P_{\alpha\theta\mu\nu}=AT_{\alpha\theta\mu\nu}+BT_{\mu\nu\alpha\theta}+\\
&+\frac{C}{4}(T_{\alpha\mu\theta\nu}-T_{\theta\mu\alpha\nu}+T_{\theta\nu\alpha\mu}-T_{\alpha\nu\theta\mu})
\label{parametricconformalcurvature}
\end{eqnarray}
having the same symmetries and being traceless and also conformally covariant in $(1+3)$-dimensional spacetimes, whose usefulness will turn out in the following.

Next we will follow the notation \cite{F,FABBRI} for the standard model. In particular the fields $B_{\mu}$ and $\vec{A}_{\mu}$ are vectors having gauge transformations given by the abelian $U(1)$ and the simplest non-abelian $SU(2)$ group and with trivial conformal transformation, from which it is possible to define the gauge-covariant derivatives $D_{\alpha}$ and the Maxwell and Yang-Mills gauge curvatures
\begin{eqnarray}
&B_{\mu\nu}=\partial_{\mu}B_{\nu}-\partial_{\nu}B_{\mu}
\label{gaugecurvatureB}\\
&\vec{A}_{\mu\nu}=\partial_{\mu}\vec{A}_{\nu}-\partial_{\nu}\vec{A}_{\mu}
+g\vec{A}_{\mu}\times\vec{A}_{\nu}
\label{gaugecurvatureA}
\end{eqnarray}
antisymmetric for indices transposition and traceless for indices contraction and conformally covariant in $(1+3)$-dimensions, as expected for gauge fields.

The fermion fields are introduced as a single right-handed spinor $\psi_{R}$ and a doublet of left-handed spinors $\psi_{L}$ defined in terms of their transformation law under the same $U(1) \times SU(2)$ group and with scaling $\sigma^{-\frac{3}{2}}$ while the scalar field is a doublet of complex scalar fields $\phi$ set by its transformation law under the same $U(1) \times SU(2)$ group and with scaling $\sigma^{-1}$ as usual.
%%%%%%%%%%%%%%%%%%%%%%%%%%%%%%%%%%%%%%%%%%%%%%%%%%%%%%%%%%%%%%%%%%%%%%%%%%%%%%%%%%%%%%%%%%%%%%%%%%%
%%%%%%%%%%%%%%%%%%%%%%%%%%%%%%%%%%%%%%%%%%%%%%%%%%%%%%%%%%%%%%%%%%%%%%%%%%%%%%%%%%%%%%%%%%%%%%%%%%%
\section{Conformal Standard Model}
For the Standard Model as we know it \cite{F}, we have that the action is a scalar and gauge symmetric under the $U(1)\times SU(2)$ group and not conformally invariant as the Lagrangian has terms that do not scale by the $\sigma^{-4}$ factor, but instead they scale by the $\sigma^{-2}$ factor; these terms are given by the Ricci curvature $G$ necessary for the gravitational dynamics and the scalar quadratic potential $\phi^{2}$ essential to bring the trivial vacuum in a non-stable configuration that is supposed to eventually move toward a stable configuration with non-trivial vacuum.

For a conformal version of the Standard Model instead \cite{FABBRI}, the action must be a scalar and gauge symmetric under the $U(1)\times SU(2)$ group and also conformally invariant with a Lagrangian that scales by the $\sigma^{-4}$ factor; consequently we need to have, on the one hand, terms like the square of the $T_{\alpha\theta\mu\nu}$ tensor used to determine the gravitational dynamics, while, on the other hand, the product between the curvature $M$ and the quadratic potential $\phi^{\dagger}\phi=\phi^{2}$ may be used as a potential for the conformal-gauge dynamical symmetry breaking.

Actually the approach enjoys a particular elegance that can be appreciated by noticing the following fact: under a global conformal transformation both the scalar dynamical term $D_{\rho}\phi^{\dagger} D^{\rho}\phi$ and the scalar potential term $\phi^{2}M$ scale by the correct $\sigma^{-4}$ factor, although under a local more general conformal transformation these two terms will be accompanied by extra pieces that would spoil the invariance, unless a proper fine-tuning is chosen so to have them all cancelling exactly, yielding a conformally invariant scalar action; on the other hand however, general conformal transformation in presence of metric and torsion widen the range of possibilities because beyond the usual term $\phi^{2}M$ there are additional terms like $\phi^{2}Q_{\alpha}Q^{\alpha}$ as well as $D_{\nu}\phi^{2}Q^{\nu}$ that may be taken into account beside the dynamical term $D_{\rho}\phi^{\dagger} D^{\rho}\phi$ and, since expressions with derivatives of the scalar and torsion such as for instance 
$D_{\rho}\phi+\frac{1}{3q}Q_{\rho}\phi$ are conformally covariant, then it is possible to restore the metric-torsional conformal invariance of the scalar action as a whole. Under this point of view, all potentials of conformal-gauge symmetry breaking are not just added for generality, but because such scalar potentials beside the scalar dynamical term are necessary to maintain the conformal-gauge symmetry before its breakdown: after the most general scalar action is found, the total action is given in the following form
\begin{eqnarray}
\nonumber
&S_{\rm{SM}}=\int[T^{\alpha\theta\mu\nu}P_{\alpha\theta\mu\nu}
-\frac{1}{4}B^{\mu\nu}B_{\mu\nu}-\frac{1}{4}\vec{A}^{\mu\nu}\cdot\vec{A}_{\mu\nu}+\\
\nonumber
&+\frac{i}{2}\left(\overline{\psi}_{R}\gamma^{\mu}D_{\mu}\psi_{R}-
D_{\mu}\overline{\psi}_{R}\gamma^{\mu}\psi_{R}\right)
+\frac{i}{2}\left(\overline{\psi}_{L}\gamma^{\mu}D_{\mu}\psi_{L}-
D_{\mu}\overline{\psi}_{L}\gamma^{\mu}\psi_{L}\right)+\\
\nonumber
&+D_{\rho}\phi^{\dagger}D^{\rho}\phi+
\left(\frac{1-6k(1-q)}{3q}\right)D_{\nu}\phi^{2}Q^{\nu}
+\left(\frac{1-6k(1-q)(1+2q)}{9q^{2}}\right)\phi^{2}Q_{\alpha}Q^{\alpha}+\\
&+k\phi^{2}M-Y\left(\overline{\psi}_{R}\phi^{\dagger}\psi_{L}
+\overline{\psi}_{L}\phi\psi_{R}\right)-\frac{\lambda}{8}\phi^{4}]\sqrt{|g|}dV
\label{action}
\end{eqnarray}
in terms of the $k$, $\lambda$ and $Y$ parameters and such that under the most general coordinate $U(1)\times SU(2)$ conformal transformation it is invariant; we vary this action with respect to the spin-connection and the vierbein taking into account that these variations are transferred through (\ref{spin-connection}) and (\ref{curvature}) onto the variations of the torsion and curvature according to the identities given by the following formulas $\delta Q^{i}_{\phantom{i}\rho\alpha}\!=\!-\left(D_{\rho}\delta e^{i}_{\alpha}
-D_{\alpha}\delta e^{i}_{\rho}-\delta e^{i}_{\sigma}Q^{\sigma}_{\phantom{i}\rho\alpha}\right)
+\left(\delta \omega^{i}_{\phantom{i}p\alpha}e_{\rho}^{p}
-\delta \omega^{i}_{\phantom{i}p\rho}e_{\alpha}^{p}\right)$ together with $\delta G^{i}_{\phantom{i}j\mu\nu}=\left(D_{\mu}\delta \omega^{i}_{j\nu}-D_{\nu}\delta \omega^{i}_{j\mu}
-\delta \omega^{i}_{j\rho}Q^{\rho}_{\phantom{\rho}\mu\nu}\right)$ showing in particular that the variation of the curvature does not depend on the variation of the vierbeins, then for the variation of the action with respect to the gauge fields we use formulas given in (\ref{gaugecurvatureB}-\ref{gaugecurvatureA}), and the variation of the spinor and the scalar fields is straightforward, so that variation with respect to the spin-connection and vierbeins gives field equations for the spin and energy densities according to
\begin{eqnarray}
\nonumber
&4[(\frac{1-q}{3q})(\frac{1}{2}Q_{\sigma\rho\beta}g^{\mu[\alpha}P^{\theta]\sigma\rho\beta}
-Q_{\rho}P^{\rho[\alpha\theta]\mu})+\\
&+D_{\rho}P^{\alpha\theta\mu\rho}+Q_{\rho}P^{\alpha\theta\mu\rho}
-\frac{1}{2}Q^{\mu}_{\phantom{\mu}\rho\beta}P^{\alpha\theta\rho\beta}]
=S^{\mu\alpha\theta}\\
\nonumber
&2[(\frac{1-q}{3q})
(Q_{\nu}(2P^{\mu\rho\alpha\nu}Q_{\rho}
-g^{\mu\alpha}P^{\nu\theta\rho\sigma}Q_{\theta\rho\sigma}
-P^{\mu\nu\rho\sigma}Q^{\alpha}_{\phantom{\alpha}\rho\sigma})+\\
\nonumber
&D_{\nu}(2P^{\mu\rho\alpha\nu}Q_{\rho}
-g^{\mu\alpha}P^{\nu\theta\rho\sigma}Q_{\theta\rho\sigma}
+g^{\mu\nu}P^{\alpha\theta\rho\sigma}Q_{\theta\rho\sigma}))+\\
\nonumber
&+P^{\theta\sigma\rho\alpha}T_{\theta\sigma\rho}^{\phantom{\theta\sigma\rho}\mu}
-\frac{1}{4}g^{\alpha\mu}P^{\theta\sigma\rho\beta}T_{\theta\sigma\rho\beta}
+P^{\mu\sigma\alpha\rho}M_{\sigma\rho}]+\\
&+\frac{1}{2}[(B^{\alpha\rho}B^{\mu}_{\phantom{\mu}\rho}-\frac{1}{4}B^{2}g^{\alpha\mu})
+(\vec{A}^{\alpha\rho}\cdot\vec{A}^{\mu}_{\phantom{\mu}\rho}-\frac{1}{4}A^{2}g^{\alpha\mu})]
=\frac{1}{2}T^{\alpha\mu}
\end{eqnarray}
whereas the variation with respect to the pair of gauge potentials gives the couple of field equations for the two currents according to
\begin{eqnarray}
&D_{\rho}B^{\rho\mu}+Q_{\rho}B^{\rho\mu}
+\frac{1}{2}Q^{\mu\beta\rho}B_{\beta\rho}=J^{\mu}\\
&D_{\rho}\vec{A}^{\rho\mu}+Q_{\rho}\vec{A}^{\rho\mu}
+\frac{1}{2}Q^{\mu\beta\rho}\vec{A}_{\beta\rho}=\vec{J}^{\mu}
\end{eqnarray}
where the spin and energy densities are given by
\begin{eqnarray}
\nonumber
&S^{\mu\alpha\theta}=\frac{i}{4}\overline{\psi}_{R}\{\gamma^{\mu},\sigma^{\alpha\theta}\}\psi_{R}
+\frac{i}{4}\overline{\psi}_{L}\{\gamma^{\mu},\sigma^{\alpha\theta}\}\psi_{L}+\\
\nonumber
&+\left(\frac{6k-1}{6q}\right)\left(D^{\theta}\phi^{2}g^{\alpha\mu}
-D^{\alpha}\phi^{2}g^{\theta\mu}\right)+\\
&+\left(\frac{1-6k-3kq^{2}}{9q^{2}}\right)\phi^{2}\left(Q^{\alpha}g^{\theta\mu}
-Q^{\theta}g^{\alpha\mu}\right)-k\phi^{2}Q^{\mu\alpha\theta}\\
\nonumber
&T^{\alpha\mu}=\frac{i}{2}\left(\overline{\psi}_{R}\gamma^{\alpha}D^{\mu}\psi_{R}-
D^{\mu}\overline{\psi}_{R}\gamma^{\alpha}\psi_{R}\right)
+\frac{i}{2}\left(\overline{\psi}_{L}\gamma^{\alpha}D^{\mu}\psi_{L}-
D^{\mu}\overline{\psi}_{L}\gamma^{\alpha}\psi_{L}\right)+\\
\nonumber
&+\left(D^{\alpha}\phi^{\dagger}D^{\mu}\phi+D^{\mu}\phi^{\dagger}D^{\alpha}\phi
-g^{\alpha\mu}D_{\rho}\phi^{\dagger}D^{\rho}\phi\right)-\\
\nonumber
&-\left(\frac{1-6k(1-q)}{3q}\right)
\left(D^{\mu}D^{\alpha}\phi^{2}-D^{2}\phi^{2}g^{\alpha\mu}-Q^{\alpha}D^{\mu}\phi^{2}\right)-\\
\nonumber
&-2\left(\frac{1-6k(1-q^{2})}{9q^{2}}\right)
\left(D^{\mu}\phi^{2}Q^{\alpha}-g^{\alpha\mu}D_{\nu}\phi^{2}Q^{\nu}
+\phi^{2}D^{\mu}Q^{\alpha}-g^{\alpha\mu}\phi^{2}D_{\rho}Q^{\rho}\right)+\\
\nonumber
&+\left(\frac{1-6k(1-q)}{9q^{2}}\right)g^{\alpha\mu}\phi^{2}Q^{\nu}Q_{\nu}
-2k\left(\frac{1-q}{3q}\right)
\phi^{2}\left(Q^{\alpha}Q^{\mu}+Q^{\alpha\mu\theta}Q_{\theta}\right)+\\
&+2k\phi^{2}\left(M^{\alpha\mu}-\frac{1}{2}g^{\alpha\mu}M\right)
+\frac{\lambda}{8}g^{\alpha\mu}\phi^{4}
\end{eqnarray}
and the currents are given by
\begin{eqnarray}
&J^{\mu}=-g'\overline{\psi}_{R}\gamma^{\mu}\psi_{R}
-\frac{g'}{2}\overline{\psi}_{L}\gamma^{\mu}\psi_{L}
-\frac{ig'}{2}\left(D^{\mu}\phi^{\dagger}\phi-\phi^{\dagger}D^{\mu}\phi\right)\\
&\vec{J}^{\mu}=-\frac{g}{2}\overline{\psi}_{L}\gamma^{\mu}\vec{\sigma}\psi_{L}
+\frac{ig}{2}\left(D^{\mu}\phi^{\dagger}\vec{\sigma}\phi
-\phi^{\dagger}\vec{\sigma}D^{\mu}\phi\right)
\end{eqnarray}
while varying with respect to the spinor we get the spinorial field equations
\begin{eqnarray}
&i\gamma^{\mu}D_{\mu}\psi_{R}+\frac{i}{2}Q_{\mu}\gamma^{\mu}\psi_{R}-Y\phi^{\dagger}\psi_{L}=0
\label{Diracfieldequations1}\\
&i\gamma^{\mu}D_{\mu}\psi_{L}+\frac{i}{2}Q_{\mu}\gamma^{\mu}\psi_{L}-Y\phi\psi_{R}=0
\label{Diracfieldequations2}
\end{eqnarray}
and varying with respect to the scalar we have the scalar field equations
\begin{eqnarray}
\nonumber
&D^{2}\phi+Q^{\rho}D_{\rho}\phi+\left(\frac{1-6k(1-q)}{3q}\right)D_{\nu}Q^{\nu}\phi+\\
&+\left(\frac{-1+3q+6k-12qk+6kq^{2}}{9q^{2}}\right)Q^{\nu}Q_{\nu}\phi
-kM\phi+\frac{\lambda}{4}\phi^{2}\phi+Y\overline{\psi}_{R}\psi_{L}=0
\label{Higgsfieldequation}
\end{eqnarray}
as the system of field equations of the conformal Standard Model. Finally it is possible to see that in this set of field equations when the Dirac and scalar field equations are considered for the energy and spin and the two currents then the conserved quantities satisfy the following conservation laws arising from the invariance under gauge phase transformations
\begin{eqnarray}
&D_{\mu}J^{\mu}+Q_{\mu}J^{\mu}=0\\
&D_{\mu}\vec{J}^{\mu}+Q_{\mu}\vec{J}^{\mu}=0
\end{eqnarray}
the spacetime frame transformations
\begin{eqnarray}
&D_{\mu}T^{\mu\rho}+Q_{\mu}T^{\mu\rho}-T_{\mu\sigma}Q^{\sigma\mu\rho}
+S_{\theta\mu\sigma}G^{\sigma\mu\theta\rho}
+J_{\mu}B^{\mu\rho}+\vec{J}_{\mu}\cdot\vec{A}^{\mu\rho}=0\\
&D_{\rho}S^{\rho\mu\nu}+Q_{\rho}S^{\rho\mu\nu}
+\frac{1}{2}T^{[\mu\nu]}=0
\end{eqnarray}
and spacetime conformal scaling 
\begin{eqnarray}
&(1-q)(D_{\mu}S_{\nu}^{\phantom{\nu}\nu\mu}+Q_{\mu}S_{\nu}^{\phantom{\nu}\nu\mu})
+\frac{1}{2}T_{\mu}^{\phantom{\mu}\mu}=0
\end{eqnarray}
and for which the Jacobi-Bianchi identities are verified identically.

We have to notice two important issues: the first is that in the energy density of the spinor field there is no explicit torsional contribution, and this is due to the fact that the variation with respect to the vierbein of the spinor-covariant derivative of the spinor field vanishes identically therefore developing no term that need to be integrated by parts; the second is that the scalar field contributes to the spin density, and this is due to the coupling between torsion and scalar fields, forced by conformal invariance. Notice also that the gauge field equations for the currents are the same in the non-conformal and the conformal version of the standard model. A further step consists in decomposing the full connection into the torsionless connection plus torsional contributions, so that the torsionless connection known as Levi-Civita connection gives covariant derivatives $\nabla_{\mu}$ and curvature tensors $R_{\alpha\beta\mu\nu}$ whose irreducible part $C_{\alpha\beta\mu\nu}$ is the Weyl conformal curvature while torsion itself in its three irreducible parts is
\begin{eqnarray}
&Q_{\alpha\mu\nu}=\frac{1}{3}\left(g_{\alpha\mu}Q_{\nu}-g_{\alpha\nu}Q_{\mu}\right)
+\varepsilon_{\alpha\mu\nu\sigma}V^{\sigma}+T_{\alpha\mu\nu}
\end{eqnarray}
where $T_{\alpha\mu\nu}$ is the non-completely antisymmetric irreducible part: then in the spinorial field equations (\ref{Diracfieldequations1}-\ref{Diracfieldequations2}) because of the extra term, torsion trace contributions cancel exactly leaving only the torsional dual axial contributions
\begin{eqnarray}
&i\gamma^{\mu}\nabla_{\mu}\psi_{R}-\frac{3}{4}V_{\mu}\gamma^{\mu}\psi_{R}
-Y\phi^{\dagger}\psi_{L}=0\\
&i\gamma^{\mu}\nabla_{\mu}\psi_{L}+\frac{3}{4}V_{\mu}\gamma^{\mu}\psi_{L}
-Y\phi\psi_{R}=0
\end{eqnarray}
as usual and where the different sign highlights the chirality of this type of self-interactions, while in the scalar field equations (\ref{Higgsfieldequation}) we have that
\begin{eqnarray}
\nonumber
&\nabla^{2}\phi+\left(\frac{1-6k}{3q}\right)\nabla_{\nu}Q^{\nu}\phi-\left(\frac{1-6k}{9q^{2}}\right)Q^{\nu}Q_{\nu}\phi-\frac{3k}{2}V^{\nu}V_{\nu}\phi-\\
&-\frac{k}{2}T^{\nu\pi\alpha}T_{\nu\pi\alpha}\phi-kR\phi
+\frac{\lambda}{4}\phi^{2}\phi+Y\overline{\psi}_{R}\psi_{L}=0
\end{eqnarray}
and where we see that the configuration $\phi\equiv0$ is not a solution in presence of spinorial fields. We have that instead if spinorial fields are present more general solutions with a non-vanishing vacuum expectation value must be sought.
%%%%%%%%%%%%%%%%%%%%%%%%%%%%%%%%%%%%%%%%%%%%%%%%%%%%%%%%%%%%%%%%%%%%%%%%%%%%%%%%%%%%%%%%%%%%%%%%%%%
\subsection{Higgs Potential for Symmetry Breaking}
The Conformal Standard Model built so far has an advantage with respect to the ordinary Standard Model \cite{F}, which consists in the fact that because of the presence of conformal gravitational degrees of freedom there is a more complicated form for the Higgs potential inducing now a dynamical symmetry breaking phenomenon: in fact after a direct calculation we have that once the vacuum expectation value for the Higgs is given by $v^{2}=\phi^{2}$ the condition for the stable stationary point for which the symmetry is broken is given by
\begin{eqnarray}
&\frac{\lambda}{4}v^{2}=\left(\frac{6k-1}{3q}\right)\nabla_{\nu}Q^{\nu}
-\left(\frac{6k-1}{9q^{2}}\right)Q^{\nu}Q_{\nu}+\frac{3k}{2}V^{\nu}V_{\nu}
+\frac{k}{2}T^{\nu\pi\alpha}T_{\nu\pi\alpha}+kR
\label{Higgspotential}
\end{eqnarray}
which we are now going to discuss in some detail: notice first of all that if the constant $k$ has the value it would have in the torsionless case $k=\frac{1}{6}$ then (\ref{Higgspotential}) would reduce to the simpler expression given by
\begin{eqnarray}
&\lambda v^{2}=V^{\nu}V_{\nu}+\frac{1}{3}T^{\nu\pi\alpha}T_{\nu\pi\alpha}+\frac{2}{3}R
\label{Higgspotentialsimple}
\end{eqnarray}
in which we see that even in absence of torsion non-trivial solutions are possible whenever $2R=3\lambda v^{2}$ is satisfied, which it is in a de Sitter spacetime of negative spatial curvature, as reported for instance in \cite{FABBRI}; however, this condition relies upon the curvature of the spacetime and its purely spatial projection, which is certainly considerable in cosmology but it might turn out to be small in particle physics, and therefore we shall retain the presence of torsion in order to ensure that (\ref{Higgspotentialsimple}) has non-trivial solutions even in a context in which the metric is approximately flat. Actually, even if the metric were not flat and curvature large, the study of torsional effects for the symmetry breaking would still be of some interest: if however the metric is flat or curvature small, we would have
\begin{eqnarray}
&\lambda v^{2}=V^{\nu}V_{\nu}+\frac{1}{3}T^{\nu\pi\alpha}T_{\nu\pi\alpha}
\label{Higgspotentialsimpleover}
\end{eqnarray}
and the presence of torsion would allow new ways to provide the dynamical symmetry breaking mechanism in the conformal Standard Model.
%%%%%%%%%%%%%%%%%%%%%%%%%%%%%%%%%%%%%%%%%%%%%%%%%%%%%%%%%%%%%%%%%%%%%%%%%%%%%%%%%%%%%%%%%%%%%%%%%%%
\subsection{Generation of Mass and Cosmological Constant}
As it is widely known, after the vacuum expectation value is gotten by the Higgs field a symmetry breaking occurs because the new ground state of the Higgs field is no longer invariant, and the new Higgs field is seen as a fluctuation over the special ground state, therefore producing two mechanisms: on the one hand, there is generation of the masses of the particles that couple to the Higgs, taking place in two ways: both as a transfer of degrees of freedom from the Higgs to the massless bosons which then become massive bosons, and as the result of the presence of the potential of interaction between Higgs and fermions and of self-interaction of the Higgs with itself; on the other hand, there is the appearance of the cosmological constant, again as the result of the presence of the same potential of self-interaction of the Higgs with itself. In the following, we shall not take into account the mechanism of the generation of the masses of the bosons, thoroughly discussed in the literature; we will instead focus on the generation of the masses of the fermion and Higgs field and of the cosmological constant, whose origin is due to the presence of the Yukawa and Higgs potentials.

After a straightforward calculation, it is easy to see that the values of the mass of the fermion and the Higgs and also the cosmological constant
\begin{eqnarray}
&m_{\rm{fermion}}=Yv\label{fermion}\ \ \ \ \ \ \ \ 
m_{\rm{Higgs}}^{2}=\frac{\lambda v^{2}}{2}\label{Higgs}\ \ \ \ \ \ \ \ \ \ \ \ \ \ \ \ 
\Lambda=\frac{\lambda v^{4}}{16}\label{cosmological}
\end{eqnarray}
are given by the Yukawa coupling $Y$ and the Higgs parameter $\lambda$ in terms of the vacuum expectation value $v$ of the Higgs field itself. Notice that as the Yukawa coupling is unknown the knowledge of the fermion mass does not give any clue about $v$ whose value of about $350 \ \rm{GeV}$ is determined when the low-energy limit of the fermion scattering is compared to the effective Fermi scattering, and this value is used to evaluate from the Higgs mass the cosmological constant.

In fact by combining the two definitions above we have that
\begin{eqnarray}
&\Lambda=\left(\frac{m_{\rm{Higgs}}v}{2\sqrt{2}}\right)^{2}
\end{eqnarray}
which with the vacuum expectation value of about $350 \ \rm{GeV}$ and the Higgs mass at least of the order of magnitude of $10^{2} \ \rm{GeV}$ gives a cosmological constant at least of order of magnitude of $10^{8} \ \rm{GeV}^{4}$ which is far from the upper limit of the order of magnitude of $10^{-46} \ \rm{GeV}^{4}$ as astrophysical experiments tell.

This situation, in which the ground state of the Higgs field gives reasonable masses only at the price of having a largely wrong cosmological constant, gives rise to what is usually known as the cosmological constant problem: this problem may be solved by a mechanism for which the vacuum expectation value of the Higgs field diminishes as we move from particle physics to cosmology, and this is precisely what happens in this approach. Nevertheless, the cosmological problem might be circumvented if we think that conformal Weyl gravity and standard Einstein gravity are quite different in their structure, and correspondingly there are two different treatments for the cosmological constant problem, so that what appears to be a wrong prediction in standard cosmologies might have no conflict with observations in conformal cosmologies \cite{Mannheim:2009qi}.
%%%%%%%%%%%%%%%%%%%%%%%%%%%%%%%%%%%%%%%%%%%%%%%%%%%%%%%%%%%%%%%%%%%%%%%%%%%%%%%%%%%%%%%%%%%%%%%%%%%
%%%%%%%%%%%%%%%%%%%%%%%%%%%%%%%%%%%%%%%%%%%%%%%%%%%%%%%%%%%%%%%%%%%%%%%%%%%%%%%%%%%%%%%%%%%%%%%%%%%
\section*{Conclusion}
In the present paper, we have considered the fully endowed metric-torsional conformal Standard Model writing the most generally invariant action and deriving the field equations: we have isolated the Higgs sector determining the ground state for the stable stationary potential that determines dynamical symmetry breaking generating masses and cosmological constant; we have discussed the vacuum expectation value in some cases, calculating the generated masses and the cosmological constant; we have discussed how the cosmological constant problem arises. The results we have found indicate that the cosmological problem may be solved by a model in which the vacuum expectation value tends to vanish as we move the energy scale from particle physics toward cosmological systems: we have discussed how this is impossible where the Higgs vacuum expectation value is constant as in the Standard Model, possible but might need fine-tunings where the Higgs vacuum expectation value is a parameter as in the metric conformal Standard Model, possible in a dynamical way where the Higgs vacuum expectation value is a field as in the metric-torsional conformal Standard Model; we also reported that in conformal gravity the possibility to exploit scale transformation may be use to control the value of the cosmological constant, and the cosmological constant problem might have consequences less severe than those it has in the non-conformal Standard Model, and the cosmological constant problem might not even arise \cite{Mannheim:2009qi}. It is then easy to argue that, if beside the metric also torsion is considered and if gravity receives a conformal treatment, when discussing the Standard Model, then both advantages presented here and in \cite{Mannheim:2009qi} can be used simultaneously, in order either to solve or avoid the cosmological constant problem. The discussion presented is therefore essential to address one of the most important problems in physics.
%%%%%%%%%%%%%%%%%%%%%%%%%%%%%%%%%%%%%%%%%%%%%%%%%%%%%%%%%%%%%%%%%%%%%%%%%%%%%%%%%%%%%%%%%%%%%%%%%%%
%%%%%%%%%%%%%%%%%%%%%%%%%%%%%%%%%%%%%%%%%%%%%%%%%%%%%%%%%%%%%%%%%%%%%%%%%%%%%%%%%%%%%%%%%%%%%%%%%%%

%%%%%%%%%%%%%%%%%%%%%%%%%%%%%%%%%%%%%%%%%%%%%%%%%%%%%%%%%%%%%%%%%%%%%%%%%%%%%%%%%%%%%%%%%%%%%%%%%%%
\end{document}